\documentclass[a4paper,11pt]{article}

\usepackage{jinstpub} 

\usepackage{lineno}
\usepackage{subfig}

\title{\boldmath Measurement of Photons Emitted by High-Energy Charged Particles as Background in Single-Photon Resolving Image Sensors}

\author[a,b,1]{Guillermo Fernandez Moroni \note{Corresponding author.}}
\author[c,d]{Fernando Chierchie}
\author[e]{Lucas Giardino}
\author[a]{Javier Tiffenberg}
\author[a,b]{Juan Estrada}

\affiliation[a]{Fermi National Accelerator Laboratory, P.O. Box 500, Batavia, Il 60510, USA}
\affiliation[b]{Department of Astronomy and Astrophysics, University of Chicago, 5640 South Ellis Avenue, Chicago, IL, 60637, USA}
\affiliation[c]{Departamento de Ingenier\'ia El\'ectrica y de Computadoras (DIEC), Universidad Nacional del Sur (UNS), Bah\'ia Blanca, 8000, Argentina}
\affiliation[d]{Instituto de Inv. en Ing. El\'ectrica ``Alfredo Desages'' (IIIE), CONICET, Bah\'ia Blanca, 8000, Argentina}
\affiliation[e]{Universidad de Buenos Aires,\\Buenos Aires, Argentina.}

\emailAdd{gfmoroni@fnal.gov}

\abstract{This work introduces an advanced technique optimized for detecting photons generated by charged particles, leveraging Skipper-CCD image sensors. By analyzing background sources and detection efficiencies, the technique achieves strong agreement between experimental results and Cherenkov-based simulations. It also provides a robust framework for investigating secondary photon production in environments with high fluxes of ionizing particles, such as those anticipated in space-based astronomical instruments. These secondary photons present a critical challenge as background noise for next-generation single-photon resolving imagers used to study faint celestial objects. Furthermore, the method exhibits significant potential for broader applications, including exploring photon generation in various substrate materials and examining their transport through multiple interfaces.}

\begin{document}
\maketitle
\flushbottom

\section{Introduction}
\label{sec:intro}

Recent silicon imagers able to detect single electrons or single visible photon \cite{quanta, Tiffenberg2017, Hynecek2003} enabled the observation of signals with photon rates orders of magnitude below the predecessor technologies by increasing the signal-to-noise ratio for single collected carriers in the pixel. The readout noise amplitude on these devices is much lower than the equivalent signal for one collected optical photon, preventing the electronic noise from adding a significant background signal to the measurement. At the same time, the ability to count single electrons enabled the exploration of other sources of errors in the sensor that were completely hidden by higher readout noise. One example is the discrimination of the different intrinsic mechanisms that produce free carriers in the active volume obtained using the Skipper Charge Coupled Device (or Skipper-CCD) \cite{barak2022sensei}. Recent results on similar devices reported extremely low rates \cite{sensei_2025} when the sensor is operated deep underground. 

The Skipper-CCD has been identified as a candidate technology for astronomical experiments thanks to its single-photon detection capability. In particular, the Habitable World Observatory (HWO) aims to search for habitable exoplanets in future space missions \cite{NAP26141}. Meager rates of photons are dispersed over a large detector area to evaluate their spectral content. The high quantum efficiency in the near-infrared regime due to thick Skipper-CCD combined with its single photon resolution provides a powerful tool to search for water vapor signatures in the exoplanet's atmosphere \cite{Stark_2019}. However, the harsh environment due to the high flux of high energy radiation expected \cite{RauscherNASA2019} represents a challenge for the photon detection performance of the sensor. An example of detection degradation is the reduction of the sensor's active area by large ionization tracks produced by high-energy charged particles crossing the sensor. This paper will focus on a less explored background signal, which is the secondary production of photons by the high energy particles through the Cherenkov process \cite{Rouven_2020}. These secondary photons will be background signals indistinguishable from the exoplanet photons, and they could be produced by the sensor's materials or by those surrounding it. A simulation of these effects for the Skipper-CCD in space has been presented in \cite{gaido_2024}.

In this paper, we present a dedicated experimental study to evaluate the production of secondary photons by the passage of charged particles through thick, fully depleted Skipper-CCDs. This study could serve as a test bench for evaluating the performance of detectors and surrounding materials for the production of light by high-energy particles. In addition, for the data collection, we use a novel smart-readout technique  \cite{smart_prl} to optimize the sensor operation by enhancing areas surrounding high-energy events, which are the most likely ones to present single-electron depositions from photons. The technique reduces unnecessary exposure to other background sources affecting the measurement. We also compare the measurements with the expected rate from theoretical models and identify the primary light production mechanism.

\section{Description of the experiment}

\begin{figure}[htbp]
\centering
\includegraphics[width=0.8\textwidth]{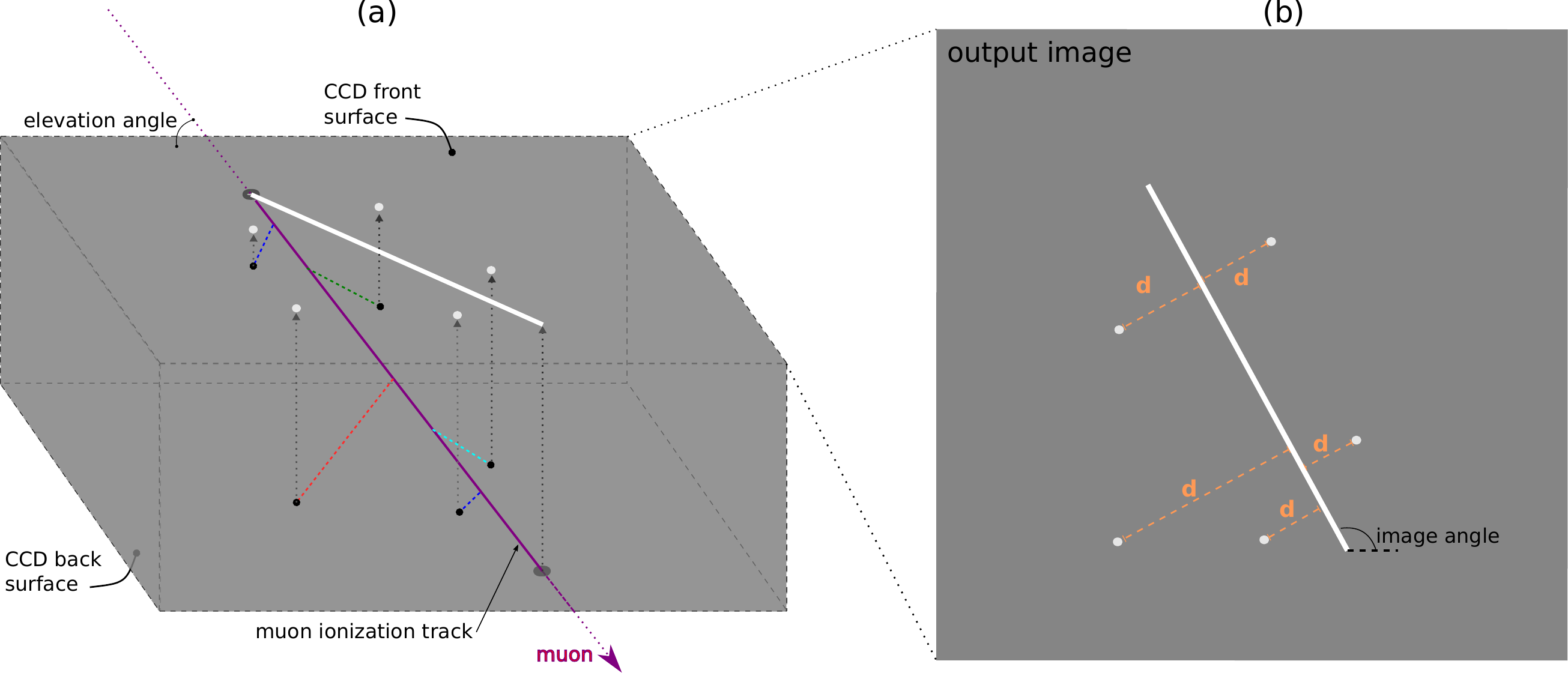}
\caption{Proposed methodology for a direct measurement of the light produced by the charged particle traveling through the sensor and surrounding materials.}
\label{fig:method}
\end{figure}

A description of the generation and detection of photons produced by the transport of charged particles through the sensor volume is presented in \ref{fig:method}(a). While the particle (in this case, a muon) transverses the sensor material, it produces ionization, represented as the purple line. The electric field in the depleted silicon drifts the free carriers toward the potential well of the pixels on the front side of the sensor. The dashed gray arrows represent this transport. The resultant white trace is the final ionization signature after its collection by the pixels. While the particle travels through the materials, it could also produce photons by several processes such as Cherenkov, Bremsstrahlung, etc. \cite{Rouven_2020}. As the Cherenkov mechanism has the most significant production rate, it was selected to compare it with the measured experimental rate. 

In the Cherenkov process, the photons are produced at a characteristic angle with respect to the direction of the traveling particle but with uniform rotation probability around it. The energy of the photons has a continuous distribution \cite{Rouven_2020}. Figure \ref{fig:method}(a) shows, in different colors, examples of trajectories traveled by different photons. The dashed line indicates the traveled path in the silicon before they are absorbed. As visually indicated by the colors, photons in the blue part of the optical spectrum have a small penetration length and are absorbed close to the track of the muon. On the other hand, photons with larger wavelengths (redder color) can travel further before being absorbed or escaping from the detector volume. If the photon is absorbed in the sensor volume, it produces a single electron-hole pair ionization. The potential wells of the pixels close to the front side of the device also collect the single carriers (holes for the sensor used in this work). The white dots in the figure \ref{fig:method}(a) represent the individual carriers collected. Photons absorbed in a position that projects close to the ionization signature of the muon track are collected by the same pixels, making it hard to separate the contributions.

Figure \ref{fig:method}(b) shows the corresponding two-dimensional output data of the explained processes. The depth information of the photon absorption position is lost during the collection of the carriers by the pixels, but the other dimensions are preserved. The production rate of photons is measured by counting the number of single-carrier ionizations around the main track. The distance of the single carrier to the main track is also recorded since it provides information on the energy distribution of the photons. As we will see later, other intrinsic and extrinsic sources of production of single-electron hole pairs could add a background signal to this measurement. 

The proposed experiment consists of measuring the photons produced by atmospheric muons in the sensor volume. Atmospheric muons have an average energy of 4 GeV, so they can transverse the sensor without being deflected, producing a straight line in the sensor. The information of the muon tracks, such as the image angle of the track in the image (as in Fig. \ref{fig:method}(b)) as well as the elevation angle (as in Fig. \ref{fig:method}(a) can be calculated using base trigonometry, and the known dimensions of the pixels and the thickness of the sensor. The three-dimensional track of the muon is used later to simulate the photon production.

\section{Experimental setup}

\begin{figure}[htbp]
\centering
\includegraphics[height=0.5\textwidth]{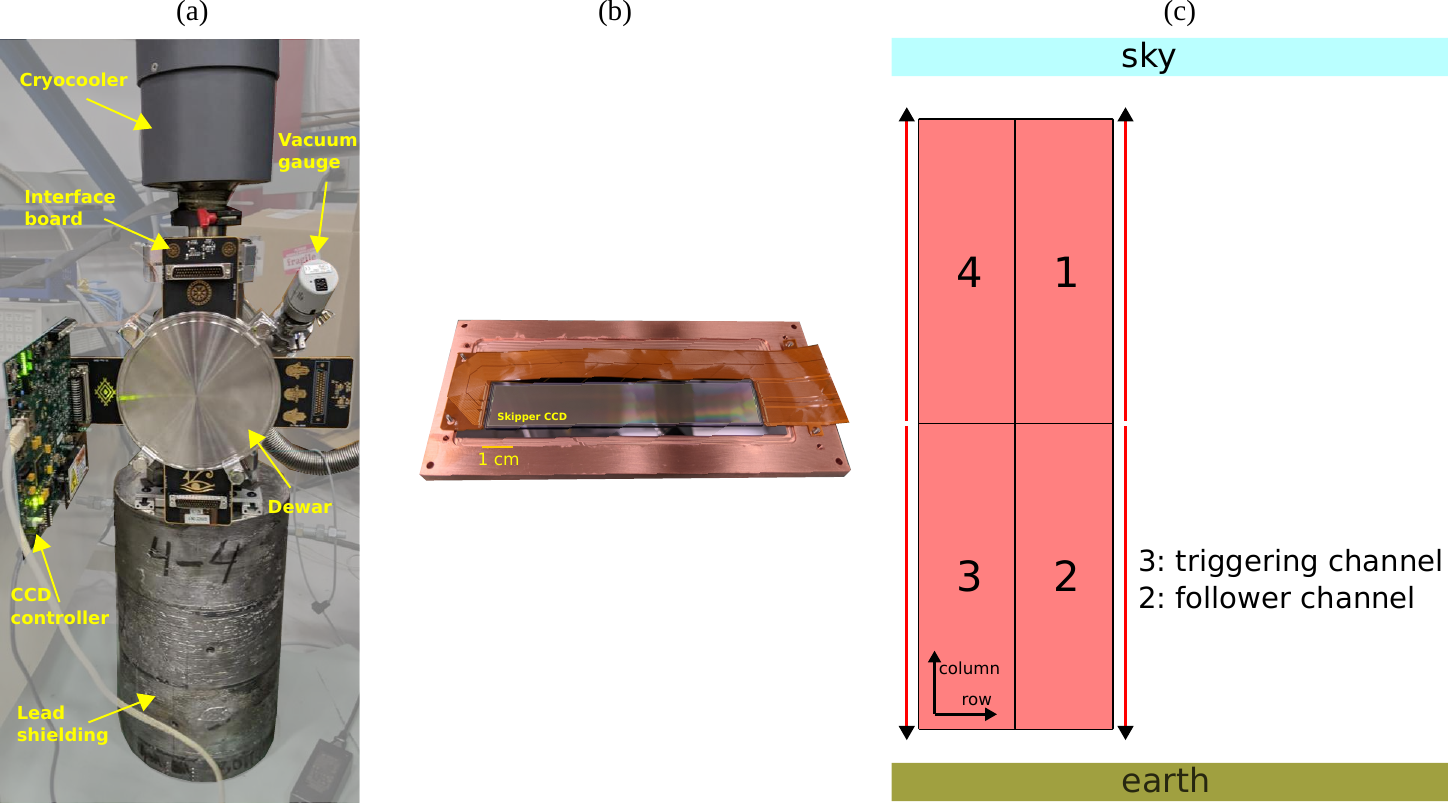}
\caption{(a) Setup used for the experiment with a short description of the main components. (b) A picture of the CCD installed on the copper tray. An extra copper plate that covers the top part of the sensor is not presented in the image. (c) Orientation of the sensor in the experimental setup.}
\label{fig:setup}
\end{figure}

Figure \ref{fig:setup} (a) shows the experimental setup where some of the main components are labeled. One Skipper-CCD (shown in Fig. \ref{fig:setup} (b)) was operated at a temperature of $140$ K using a Sunpower cryocooler \cite{Sunpower_2021}. The CCD is glued to a silicon substrate that sits on a copper tray for mechanical support as well as thermal connectivity. The CCD is placed in an extension of the dewar that fits inside a lead cylinder. There was no radio purity selection of materials inside the shield. The lead rejects other high-energy ionizing radiation, such as gamma particles that could produce additional ionization on the sensor and additional photons, which act as a background signal to the ones of interest produced by the muons. The was designed by the LBNL Microsystems Laboratory (MSL)~\cite{LBNL-MSL} and fabricated at Teledyne-DALSA. It has $6144$ columns by $1024$ rows with pixels of $15$ $\mu$m by $15$ $\mu$m with a thickness of $675$ $\mu$m operated in fully depletion mode \cite{Holland:2003}. It is read by four amplifiers, one on each corner, using a Low Threshold Acquisition (LTA) controller \cite{cancelo2021low}. Figure \ref{fig:setup} (c) depicts how the different quadrants of the sensor are oriented with respect to the earth's surface.

\begin{table}[htb]
    \centering
    \caption{Main characteristics of the CCD used in this and previous works. \label{tab:sensor}}
    \begin{tabular}{lcc}
        \hline
                        & Value             & Units        \\
        \hline
        CCD dimensions  & 6144~$\times$~886 & pixels       \\
        Number of quadrants & 4 & \\
        Pixel Size      & 15~$\times$~15    & $\mu$m$^{2}$ \\
        Thickness       & 675               & $\mu$m       \\
        CCD temperature & 140               & K            \\
        \hline
    \end{tabular}
\end{table}

\subsection{Sensor smart-readout strategy}

Each corner of the sensor has an output amplifier that measures the charge value of the pixels in each quadrant. Therefore, during readout, the charge of the pixels is transferred without distortion towards the output structure. Charge transfer on CCD devices is a very well-known process \cite{janesick_2001}. In the Skipper-CCD, Each amplifier has a floating gate sense node, which allows multiple non-destructive measurements of the pixel charge \cite{janesick_2001, skipper_2012, Tiffenberg2017}. These samples are then averaged out to get the final pixel value of the pixel. The averaging reduces the random fluctuation of the readout noise from the output transistor added to each individual measurement by the square root of the number of samples.

\begin{figure}[htbp]
\centering
\includegraphics[width=0.5\textwidth]{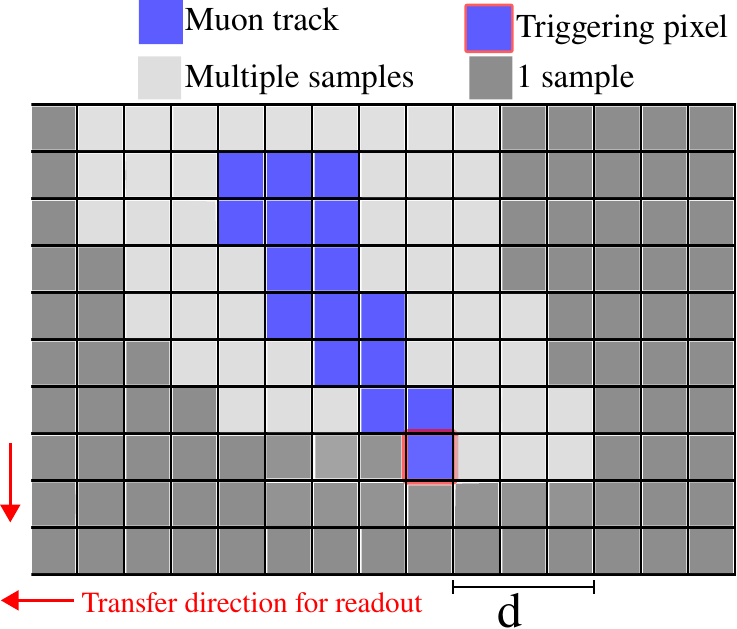}
\caption{Sketch of the readout technique implemented to inspect for light emitted by muon tracks in the sensor.}
\label{fig:smart technique}
\end{figure}

Deep sub-electron noise levels are required to detect a single carrier in the pixels. In our experimental setup, 500 samples per pixel were used with a noise level of 0.15 e$^-$ of equivalent noise. The multiple sampling readout slows down the sensor readout, increasing the exposure time to background signals, which are indistinguishable from the photon signals produced by the muons. To minimize the background signal, only the pixels surrounding the muon tracks are read out using multiple samplings with deep subelectron noise. The rest of the image was read using only one sample per amplifier to reduce the read time and minimize the exposure to light produced by other particles or sources. The smart readout technique based on the energy of interest presented in \cite{chierchie2021smartreadout} was used to identify the muon track and to switch to multiple sampling readouts to obtain single-photon resolution in these regions. Figure \ref{fig:smart technique} illustrate the technique implemented. The muons produce a continuous ionization track while traveling through the silicon. The amount of charge collected by the pixels is larger than the readout noise for one sample per pixel. Therefore, it can be used as the trigger signal to open a window of several pixels, say $d$ pixels in the same row, with multiple samples readout to search for the photon depositions. The trigger pixel is represented by a red box in Fig. \ref{fig:smart technique}. 

The readout controller remembers the column number when reading the following line, and starts a multiple readout operation $d$ pixels before it in the horizontal direction. In this case, pixels with low noise are available to search for photons at both sides of the muon track. While doing this, the readout system checks for pixels with large charge packages compatible with the ionization of the muon track. It updates the position for the following line to account for muons with oblique tracks. The multiple-sample operation ends when the readout system finds no pixel with charge compatible with the muon ionization track in a row. For the results presented here, $d=50$ pixels, which was selected to be similar to the thickness of the sensor

\begin{figure}[htbp]
\centering
\includegraphics[width=1\textwidth]{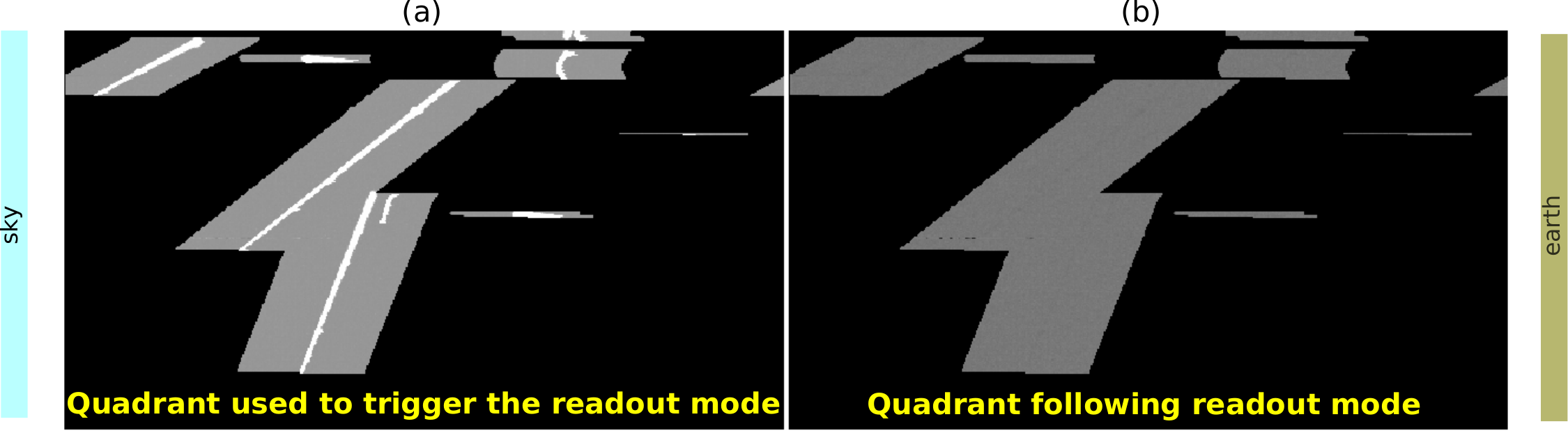}
\caption{Portion of an output image:(a) Quadrant of the sensor used to trigger the multiple sampling mode. (b) Quadrant of the sensor which follows the same readout mode as the triggering quadrant.}
\label{fig:smart technique data}
\end{figure}

Figure \ref{fig:smart technique data}(a) shows a portion of an image adquired using this procedure. The white central lines correspond to tracks in the image that trigger the multiple-sampling readout (gray areas). The black pixels only have one sample per pixel and therefore a faster readout mode. Figure \ref{fig:smart technique data}(b) shows the pixels read simultaneously from another quadrant following the same readout strategy. In this case, the multiple-sampling regions are not centered around ionization in the sensor, so these regions are used to account for the single electron rate from sources unrelated to charged-particle tracks. As we will discuss later, only the regions with multiple sampling are considered for the analysis. The percentage of all the collected data shows that only 7 \% of the image was read out with deep sub-electron noise, which is an average reduction of the readout time by a factor of 14.

\section{Data processing and simulation}

\begin{figure}[htbp]
\centering
\includegraphics[width=1\textwidth]{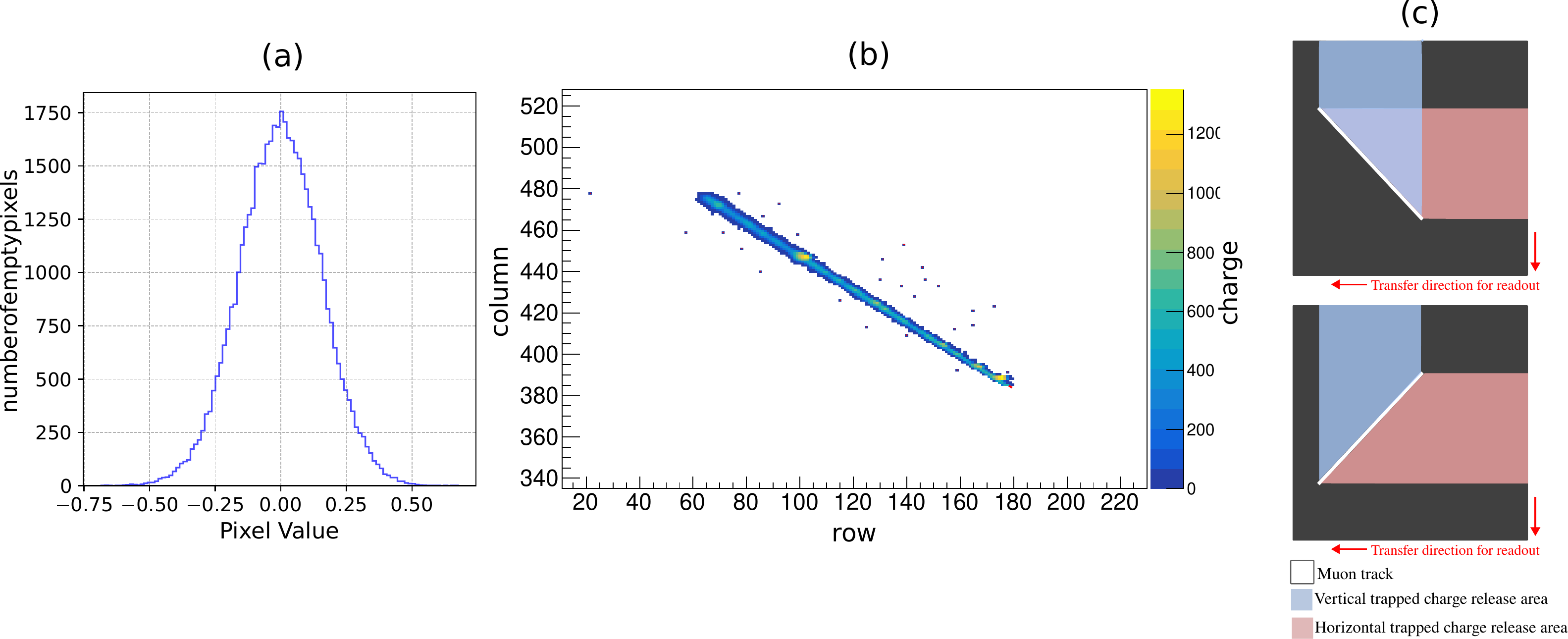}
\caption{(a)Histogram of pixels with no collected charge from a multiple sampling region. (b) Example of a system with a collected muon plus the single carrier ionization around it. (c) Sketch showing the areas with potential charge release due to charge transfer in the array of the sensor for two different directions of the muon track.}
\label{fig:data reduction}
\end{figure}

The processing focuses on the smart regions that have multiple samples per pixel. In these regions, most of the pixels have zero charge, which is helpful to estimate the offset added by the readout electronics. An offset is calculated for each region and subtracted. Figure \ref{fig:data reduction}(a) shows a histogram of pixels with zero charges after subtracting the offset. The standard deviation of the distribution is measured to evaluate the noise level of the image. With 500 samples per pixel, the typical readout noise is 0.17 e$^-$. This is selected to keep the probability of false positive probability low. Pixels with values above a threshold of 0.63e$^-$ are considered charged and are clusterized. Adjacent pixels with charge are considered from the same ionization process and recorded as part of the same event. Figure \ref{fig:data reduction}(b) shows one event extracted after the clusterization for one of the regions. The color scale gives information on the amount of charge on each pixel. Single electron ionizations surround the central muon track. The top end of the muon is thicker than the bottom end because of the extra time the carriers have to diffuse laterally before being collected by the pixel potential wells close to the front of the sensor. The other end of the muon looks narrower since the ionization is produced close to the front face. 

\subsection{Muon selection}

After the clusterization, tracks with an extension of more than 100 pixels are evaluated to check if they are compatible with a muon suitable for the proposed analysis. The selection criteria are:

\begin{itemize}
    \item The muons produce an ionization track that follows a straight line since its direction is barely modified in the silicon crystal. This feature is used to select muons over other ionizing particles. For this, the track in the image is subdivided in 30 segments. In each segment, the lateral dispersion is fitted using a Gaussian distribution. The mean of the fitted lateral distribution should not be more than 1 pixel away from a global straight-line fit of the muon. This global fit is calculated using a weighted least square fit method using the charge of the pixels in the track.

    \item We select muons tracks larger than 30 pixels in the row direction to ensure enough area to map the light production.

    \item One of the expected backgrounds is from the capture and release of single electrons by traps in the silicon \cite{janesick_2001, Cervantes-Vergara_2023}. These traps capture an electron while the charge is being transferred through them toward the output readout stage. The release time could be longer than the transfer time between pixels in the array, so the released carrier is now disconnected from the main track of the muon. We will see later that this extensive background competes with the photon signal. The image angle calculated with the global fit for each muon track can be used to select those that eliminate this contribution. Figure \ref{fig:data reduction}(c) shows the two possible main directions of the muon track in the image and the areas in the image that could be affected by this mechanism by the horizontal and vertical transfer of the pixels in the matrix. The top scenarios with muons with a negative slope are chosen for the analysis since their left are is not affected by this background signal. 

\end{itemize}

\begin{figure}
    \centering
    \includegraphics[width=0.9\columnwidth]{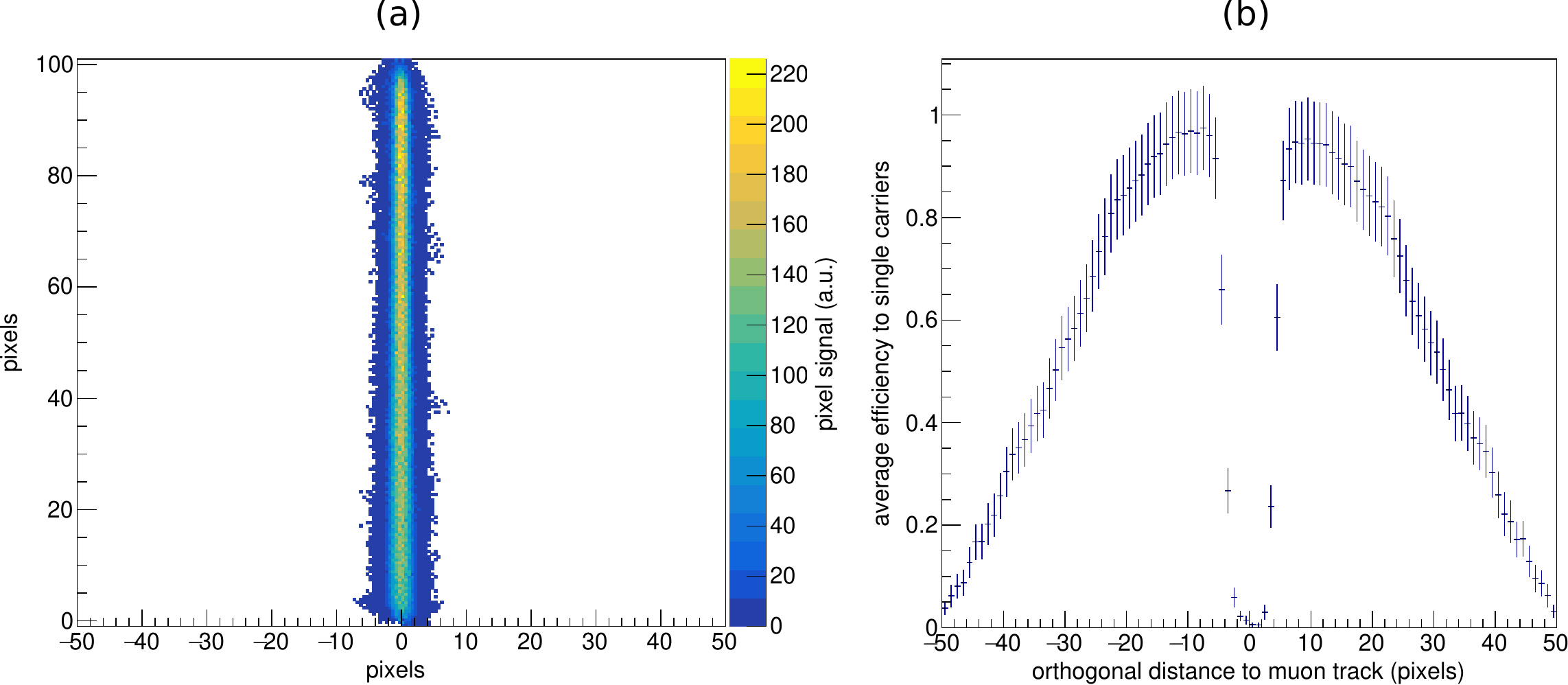}
    \caption{(a)Figure of all muons used for the analysis. The muons have been rotated and sized from their original position in the image. The figure shows that the thick part of the muon corresponds to muons crossing the rear part of the sensor while the thin end corresponds to the ionization produced at the front side. (b) Average detection efficiency for the single carrier detection measured using the selected muons.}
    \label{fig:all_muons}
\end{figure}

One hundred ninety-four events passed the selection criteria. Figure \ref{fig:all_muons}(a) shows an image where all the selected muon events are rotated to the same direction, scaled to the same length, and overlapped in the same graph. The color map shows the sum of the charge of the pixels from all the muons. The figure shows that the selection criteria work well in selecting muons and the global fit provides a good estimation of the direction of the muon. The muons are aligned by the depth of the track in the sensor, so the combined track has a larger width at the bottom than at the top of the image. Since the thickness of the detector is known, the track length on the image is used to calculate the actual 3D length of the muons' track through the material. The true track is later used to simulate the production of photons in the sensor.

\subsection{Spatial efficiency}

Some of the photons produced by the muons cannot be detected by the proposed method. There are three scenarios where the produced photons cannot be detected. The first one is when the produced photons are collected by the same pixels, or the adjacent ones, that collect the charge from the muon. In this case, the single carriers produced by the photons cannot be distinguished from the muon itself. The lateral diffusion \cite{sofo_haro_2020} widens the main track in the output image so that this effect could impact the observation of photons by a few pixels away from the centroid. The boundaries of the sensor itself give a second mechanism for losing photons. Photons can escape the volume of the detector, reducing the collection efficiency. The boundaries of the multiple sampling regions give the third effect. Photons could travel beyond the sensible region with single carrier resolution, making them indistinguishable from the sensor's readout noise. 

These mechanisms produce a reduction in the efficiency of detecting the generated photons. As the muon track direction in the sensor can be estimated, as well as the spread of the carriers before they are collected by the pixels, the efficiency for collecting single photons can be estimated. Since the muon incoming direction is random, each of them has its own efficiency, and the simulation of the detection technique accounts for that. As a reference, Figure \ref{fig:all_muons}(b) shows the average efficiency, calculated from all the selected muons, measured as a function of the perpendicular distance from the main direction of the muon in the output. For distances close to zero, when the photons are absorbed in the proximity of the muon track, the efficiency goes to zero due to the overlap with the main track. For distances above 5 pixels, the efficiency is close to 1, and the method has the highest efficiency. Beyond fifteen pixels, the efficiency starts to drop slowly, mainly due to the photons that escape the sensitive region area. For future runs, the efficiency for large distances can be improved by making the sensitive region grow following the preferred generation angle direction of the photons with respect to the muon direction. As we will see later, the results presented in this article show that the most significant signal-to-background rate is obtained for distances below 20 pixels perpendicular to the muon track.

\subsection{Simulation of the photon light production}
\label{sec:cherenkov simulation}
The production of photons is simulated using the three-dimensional trajectory of the muon transversing the sensor. The geometrical dimensions of the CCD are used to establish the boundaries of the muon interactions and the photons. The light is created in the muon track and transported in the material using the following steps:

\begin{itemize}
\item Only the Cherenkov photon production mechanism is considered since it contributes to the largest fraction of the flux \cite{Rouven_2020}.

\item The expected average number of generated photons is calculated for each selected muon track using the formalism in \cite{Rouven_2020}. For the calculation, we assumed that the muon energy is 4GeV, which is the mean energy of the muons at sea level. The average number of photons is used as the mean of a Poisson random generator to calculate the number of photons generated per muon. In our case, the simulated photon energy regime is from 1 eV to 3 eV, which represents an absorption length interval from several meters to hundreds of nanometers. More energetic photons will interact close to the muon track and are not detectable, and less energetic photons escape from the sensible area of the method.


\item The generation point of the photons is generated from a uniform distribution along the three-dimensional muon track. The photon's energy is randomly generated based on the expected energy distribution presented in \cite{Rouven_2020}. 

\item The emission angle of the photon is calculated using the characteristic angle of Cherenkov in silicon of 74.5 degrees with respect to the track of the muon. A uniform distribution in the revolution angle around the muon track generates this angle for each photon. 

\item The photon's energy is used to calculate its expected absorption length in intrinsic silicon. Then, the expectation is fed into an exponential random generator to get the absorption point in the direction fixed by the angles calculated in the previous step. 

\item If the absorption points are outside of the physical dimensions of the CCD, the photon is discarded. Those that are absorbed in the active volume of the sensor are projected towards the front surface, simulating the collection of free carriers by the pixels in the CCD. Only intrinsic silicon is considered the sensor material. No reflection effect is simulated at the sensor interfaces.

\item The projected collection of the ionization of the photons is used to build an output image together with the clusterized muon track from the data. The muon track provides the spatial area where photons can not be distinguished. The photons that do not overlap with the main track and are projected in the multiple sampling region are used to record their perpendicular distance to the muon image track.

\item This process is repeated for every muon multiple times to reduce the statistical uncertainty of the simulation.

\end{itemize}

\subsection{Single carrier background}
\label{sec:Single carrier background}
The single-electron hits that are not associated with muon tracks are measured from another channel of the sensor that follows the same readout pattern of the triggering channel. An example of this strategy is shown in Fig. \ref{fig:smart technique data}. The single-electron hits are extracted from the follower quadrant and used as a measurement of the background signal expected in the signal around the muons. The rate of this event is then used to simulate a background signal using the regions of the extracted muons from the original quadrant to weigh this rate by the efficiency of the geometric boundary of the sensitive region and the muon width. 

\section{Measured profile of photons}

\begin{figure}
    \centering
    \includegraphics[width=0.65\columnwidth]{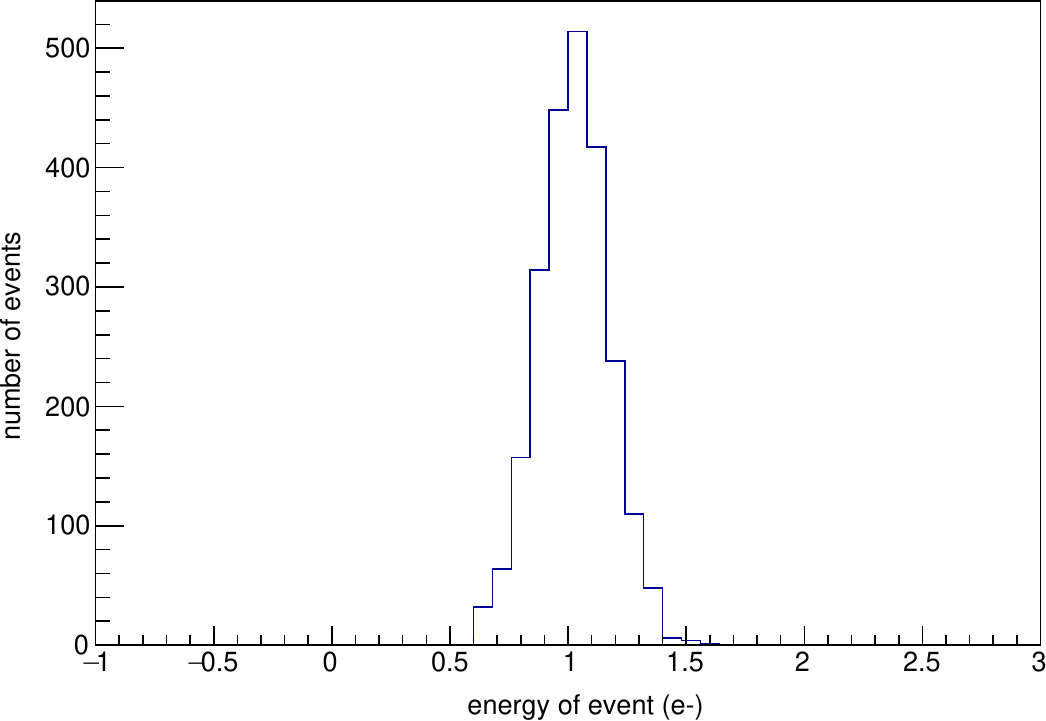}
    \caption{(a) Measured value of events with one collected carrier measured around the track of the muons.}
    \label{fig: photon energy histogram}
\end{figure}

Figure \ref{fig: photon energy histogram} shows the distribution of the charge value measured for the single carrier events detected in all sensitive regions. This distribution is centered at one electron and has a Gaussian shape, as expected from the remanent readout noise. The good symmetry of the distribution shows qualitatively that there is no significant addition of false positive pixels contributing to the single-carrier rate.

\begin{figure}
    \centering
    \includegraphics[width=0.65\columnwidth]{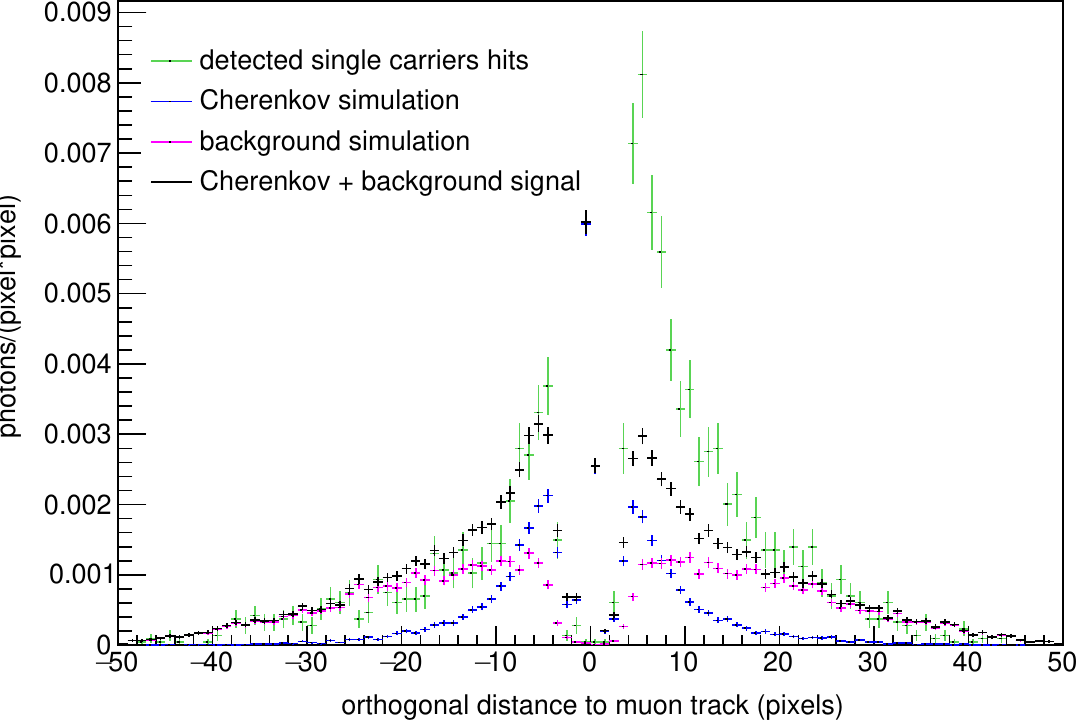}
    \caption{Distance distribution of the measured single hits around the muon tracks in the data together with the simulated distributions.}
    \label{fig: distance hit distribution}
\end{figure}

The measured rate of events as a function of the orthogonal distance to the main track is shown in Fig. \ref{fig: distance hit distribution} with green markers. The profile of events measured for each muon is added together and ultimately divided by the sum of all used muons. Each bin of the histogram has a width of one pixel of the array. Then, the histograms for all the muons are summed and normalized by the number of processed muons. The unit in the y-axis measures the rate of photons per unit length of the muon per unit length of distance to the muon track. The plot also shows the expected contribution from the Cherenkov processed from simulations with blue markers, as explained in section \ref{sec:cherenkov simulation}, and the background signal contribution with pink markers, as explained in section \ref{sec:Single carrier background}. The sum of these two contributions is shown with the black markers. Both contributions shown in the plot are calculated from each muon region as the real single electron hits explained before.

The distribution for negative distances is measured in the left region of the muons. Since only muons with negative image angles are considered for the analysis, this side of the plot has no contribution from the capture and release of carriers from the main track during charge transfer. The right side of the plot has potential contributions from both vertical and horizontal transfers of the charge in the sensor, as illustrated in Fig. \ref{fig:data reduction}(c). The distribution for negative distances shows a good agreement with the expected total contribution. For short distances, the contribution is null due to the detection inefficiency. For distances from -4 up to -20 where the contribution of the photon signal is larger than the background fluctuation and its contribution can be easily identified. In this region, the measured profile has a good agreement with the expected contribution of background plus the Cherenkov signal. For more negative distances, the contribution from the produced photons is small, and the background signal dominates the distribution. For distances greater than zero, a higher contribution of single carrier hit for distances between 4 and 20 pixels. We attribute this excess to the contribution of single carrier traps with a release constant similar to the transfer times in this experiment. 

\section{Conclusions}
\label{sec:summary}

In this work, a new optimized technique for measuring photons generated by charged particles using Skipper-CCD sensors was introduced. The technique enabled the identification and understanding of background sources and detection efficiencies, showing consistency between Cherenkov simulations and experimental results. Additionally, it proved effective in evaluating the incidence of secondary photons produced in sensors operating in environments with high fluxes of ionizing particles, such as those expected in space-based astronomical instruments. These photons represent a key background source for the next generation of single-photon resolving imagers to observe faint celestial objects. Furthermore, this methodology showed promise as a versatile tool for extending its application to the characterization of photon production in various materials used as sensor substrates and to studying photon transport across different interfaces.

\acknowledgments
Lawrence Berkeley National Laboratory (LBNL) is the developer of the fully-depleted CCD and the designer of the Skipper readout. The CCD development work was supported in part by the Director, Office of Science, of the U.S. Department of Energy under Contract No. DE-AC02-05CH11231.

This research has been partially supported by Javier Tiffenberg's and Guillermo Fernandez Moroni's DOE Early Career research programs.

\bibliography{bibs.bib} 
\bibliographystyle{unsrt}

\end{document}